\documentclass{revtex4}
\usepackage{graphicx}

\begin{document}

\title{Fast switching current detection at low\\
critical currents}

\author{J. Walter, S. Corlevi and D. B. Haviland}

\affiliation{Nanostructure Physics, Royal Institute of Technology,\\
AlbaNova, 106 91 Stockholm, Sweden}

\begin{abstract}
A pulse-and-hold technique is used to measure the switching of small critical current Josephson junctions.  %%@
This technique allows one to achieve a good binary detection and therefore measure switching probabilities.  The %%@
technique overcomes limitations on simple square pulses and allows for the measurement of junctions with critical %%@
currents of the order of $10$nA with bias pulses of the order of $100$ns. A correlation analysis of the switching %%@
events is performed to show how the switching probability depends on the wait time between repeated bias pulses.
\end{abstract}

\maketitle

\section{Introduction}

In the last several years there has been increasing interest in the quantum behavior of small capacitance Josephson %%@
junction circuits.  Progress in this field has led to quantum control experiments, which use a state preparation, %%@
evolution, and read-out, to demonstrate that the electrodynamics of these circuits behaves quantum %%@
mechanically.\cite{nakamura:CPbox:99,vion:quantronium:02,martinis:josephsonjunctionqubit:02,chiorescu:fluxqubit:03,dut%%@
y:chargequbit:03} Engineering circuits so as to isolate two energy levels and studying their quantum time evolution is %%@
interesting in the context of the long-term goal of building a solid-state, scalable quantum bit processor.  In many %%@
of these quantum control experiments, the measurement process, or the projection to a basis state and readout of that %%@
state, is achieved by quickly measuring the critical current of a Josephson junction.   In this paper, we examine this %%@
detection technique in some detail and demonstrate how it might be extended to shorter measurement times and lower %%@
critical currents.  

The quantum nature of the Josephson junction was first probed by studies of the fluctuations in the switching current, %%@
or current at which the junction jumps to the finite voltage state \cite{voss:mqt:81,devoret:quantumeffects:92}. The %%@
technique used to measure the switching current,  originally introduced by Fulton and Dunkleberger %%@
\cite{fulton:zerovoltagelifetime:73},  was based on a rapid ramping of the bias current.  With each ramping event, the %%@
junction was switched to the finite voltage state.  By fast measurement of the voltage across the junction, one could %%@
infer at what current the junction switched, and build a switching histogram, or distribution of the switching %%@
probability vs. bias current.  The Quantronics group introduced a new technique \cite{cottet:pulsereadout:02,%%@
cottet:CPTsingleshot:02} with short, square pulses ($\simeq 500$ns duration) of the current to a value near the %%@
critical current ($\simeq 0.5\mu$A) of the junction.  The pulse would cause the junction to switch with some %%@
probability.  The switching of the junction is measured in a binary way (i.e. yes or no) with each pulse. By repeating %%@
the pulse many times in a sequence, one can determine the switching probability for that particular amplitude and %%@
duration of the pulse.  This pulse method was successfully used as the readout in a quantum control %%@
experiment.\cite{vion:quantronium:02}

Both the ramping method and methods based on simple square pulses, are limited in measurement speed and accuracy by %%@
the rate of increase of the measured voltage across the junction, $dV_J/dt$.  This rate may be limited by the band %%@
width of the amplifier, or by low pass filters used to protect the junction from high frequency noise.  However, in %%@
the absence of such limitations, the junction voltage will rise at a rate $dV_J/dt=V_{bias}/R_{bias}C_{Leads}$. The %%@
speed of the measurement will then be determined by the time needed for the junction voltage to rise above the noise %%@
level, so that a switching event can be triggered.   If, however, we use a more complex shape of the voltage pulse, we %%@
can circumvent this limitation by exploiting the non-linear character of the Josephson tunnel junction I-V curve and %%@
the latching nature of the circuit.  We have developed such a method which is described below.  This method has also %%@
recently been reported by the Delft group \cite{chiorescu:fluxqubit:03}, and the Quantronics group %%@
\cite{vion:privatecomm}. Pulse methods have also been used for studying switching properties of magnetic junctions %%@
\cite{koch:magnetizationreversal:98}.

\section{Experimental method}

We have used a pulse and hold method to study the switching current of a DC SQUID.  The SQUID was chosen because the %%@
level of the critical current could be adjusted by applying a magnetic flux to the SQUID loop, allowing us to explore %%@
how the method works for a wide range of critical currents, from $270 - 30$nA with one sample.  The capacitance of the %%@
parallel combination of the two Josephson junctions is roughly $C \simeq 7.2$fF, and the normal state resistance of %%@
the SQUID (two junctions in parallel) is $R \simeq 1.15k\Omega$, giving a bare critical current value $I_{c0}= 273$nA.  %%@
When measuring the DC IV curve of the junction at zero magnetic field, we find switching at roughly $80 \% $ of this %%@
value.

\begin{figure}[!hb] 
%\epsfxsize=10cm   %width of figure - will enlarge/reduce the figures
%\epsfbox{fig3.eps}
%\figurebox{2cm}{3cm}{} %to have a box alone 
%\centerline{\epsfxsize=3in\epsfbox{schematics.eps}}   
\includegraphics[scale=0.4]{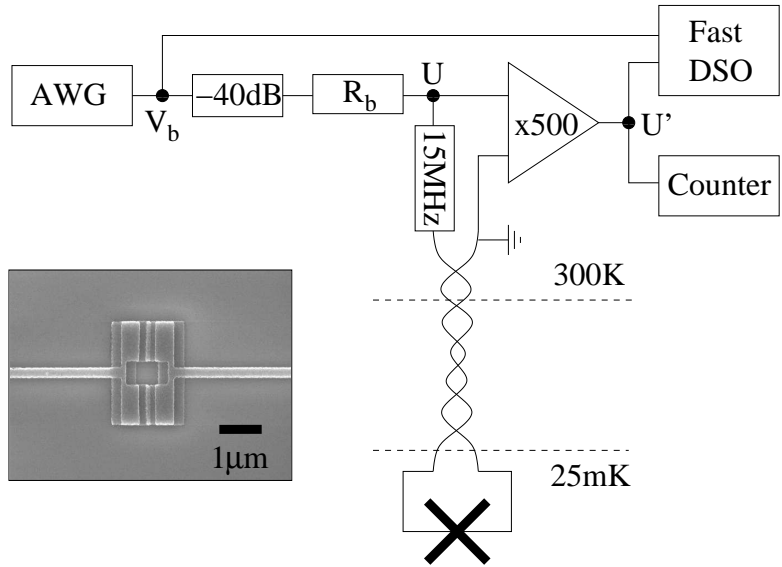}
\caption{The experimental setup and a micro graph of the sample. AWG stands for arbitrary waveform generator
and DSO stands for digital sampling oscilloscope.The picture shows a micrograph of the sample.\label{schematics}}
\end{figure}

The schematic of the measurement circuit and a picture of the sample are shown in Figure~\ref{schematics}. The SQUID %%@
sample was biased from an arbitrary waveform generator (AWG) through an attenuator, a bias resistor and a low-pass %%@
filter. The connection from room temperature to the sample in a dilution refrigerator with a base temperature of about %%@
$25$mK, was realized via twisted pair cables of constantan wire  of length $2.1$m and resistance $135\Omega$ per wire.  %%@
No special low-pass filters were implemented at low temperatures.

\begin{figure}[!ht] 
%\centerline{\epsfxsize=4.5in\epsfbox{measurements.eps}}   
\includegraphics[scale=0.4]{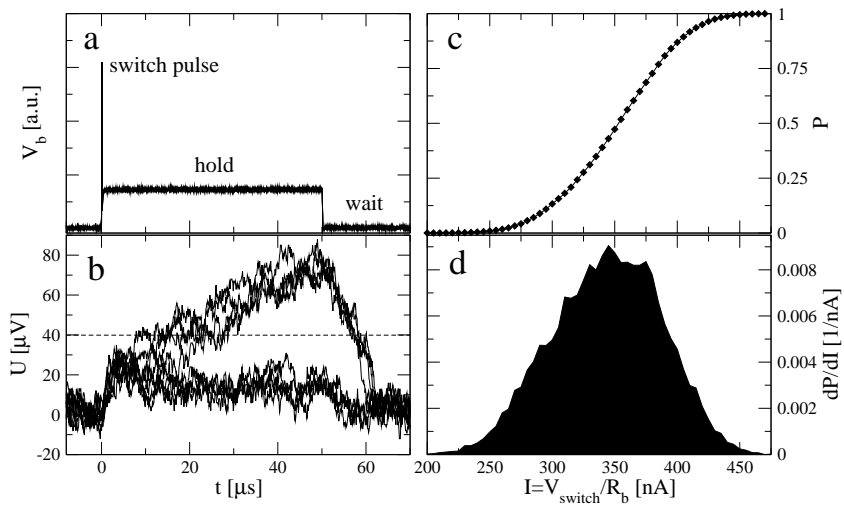}
\caption{(a) The probe pulse consisting of the switch pulse (duration $\tau _p$), hold level (duration $\tau _h$), and %%@
wait time (duration $\tau _w$) between consecutive pulses. (b) Response of the sample for $8$ identically applied %%@
pulses.(c) By successively increasing the switch pulse the switching probability P(I) is obtained. The derivative of %%@
the P(I) curve leads to the switching current distribution (d).\label{measurements}}
\end{figure}

The shape of the probe pulse as shown in fig.~\ref{measurements}a has three parts.  A very short "switch pulse" with a %%@
duration $\tau_p$ is applied, which may cause the junction to switch from the supercurrent branch to the finite %%@
voltage state.  This switch pulse can be much shorter than the time needed to charge up the capacitance at the input %%@
of the voltage amplifier, and no voltage rise is registered from this pulse alone.  The switch pulse is immediately %%@
followed by a hold level for a time $\tau_h$, so that if the junction has switched to the finite voltage state, it %%@
will not "retrap" to the zero voltage branch of the I-V curve.  The duration of the hold level is set just long enough %%@
for the junction voltage to rise above the noise level and trigger a counter to record a switching event.   When %%@
calibrating the hold level, one can check that a pulse consisting of the hold level itself, causes no switching %%@
events.  Following the hold level, a wait time $\tau_w$ is programmed in the AWG.  $\tau_w$ must be long enough for %%@
the junction to relax to the equilibrium state.  The requirements on $\tau_w$  are determined by a correlation %%@
analysis of the switching sequence of many pulses as described below.

Figure~\ref{measurements}b shows 8 typical oscilloscope traces overlaid on top of one-another, of the junction voltage %%@
$U$ measured at the top of the cryostat. By adjusting the hold level and duration, a clear separation between the %%@
switch and no-switch event can be obtained.   All the signals that rise above a trigger level of $40\mu$V are recorded %%@
as a switching event by the counter. Typically a burst of $10^4$ identical pulses is applied to the sample and the %%@
number of switching events $N$ is counted.  The switching probability is then $P=N/10^4$. It is important that a clear %%@
separation exists in the voltage response from switching and non-switching events, as shown in %%@
fig.~\ref{measurements}b. This separation is a necessary condition for a {\it binary} detector of switching and the %%@
use of event counting statistics for determining the switching probability. We find that with our measurement set-up, %%@
simple square pulses will not give this clear event separation, and ambiguities arise between switching events due to %%@
noise and those due to actual junction switching in the final moments of the square pulse duration.  

In a typical measurement, after the hold voltage level and $\tau_h$ are adjusted, the height of the switch pulse is %%@
initially chosen so that $P=0$ after $10^4$ pulses.  The amplitude of the switch pulse is then slightly increased %%@
while the other parameters of the pulse are kept constant and the burst sequence is repeated, recording a new value of %%@
$P$.  This procedure is repeated until the switching probability reaches $P=1$. In this way a $P(I)$ curve can be %%@
measured, as seen in fig.~\ref{measurements}c.  Taking the derivative of this curve we arrive at the switching current %%@
distribution shown in fig.~\ref{measurements}d.   Here, the current level is simply given by $I=V_{switch}/R_{b}$, %%@
where $V_{switch}$ is the voltage amplitude of the switch pulse as programmed in to the AWG, divided by the %%@
attenuation factor.  This method of determining the current is valid for $\tau_p>1 \mu$s.  In the experimental set-up %%@
shown in fig.~\ref{schematics} the shortest switch pulse attainable that was reasonably square was of order %%@
$\tau_p=100$ns, as measured by an oscilloscope at the bottom of the cryostat when the system was at room temperature.  %%@
Shorter pulses suffered from too much dispersion in the twisted pairs, which limited the rise and fall times of the %%@
pulses to $20$ns. Dispersion also caused a second, slow rise of order $1\mu$s, which made the actual amplitude of the %%@
pulse at the junction dependent on the pulse duration.  For this reason, it is difficult to calibrate the current %%@
level $I$ of the shortest pulses.  However, our analysis of the data relies only on a figure of merit for the %%@
detector, which is the ratio of the width of the switching distribution to the pulse height at the maximum of the %%@
distribution, $\Delta I/I$.  Thus, this unknown calibration factor for short pulses due to dispersion in the twisted %%@
pairs, falls away in the final analysis.  

%\begin{figure}[tbp] 
%\centerline{\epsfxsize=2.0in\epsfbox{ws-procs975x65/probabilityCurve.eps}}   
%\caption{(a) Switching probability as a function of the switch pulse amplitude and corresponding switching current %%@
%distribution (b).\label{scurve}}
%\end{figure}

The value of the bias resistor, $R_b$ shown in fig.~\ref{schematics} was chosen between $1$k$\Omega$ and %%@
$20$k$\Omega$, depending on the level of the switching current in the range $220$nA to $10$nA respectively.  It is %%@
desirable to make $R_b$ as small as possible for a fast rise of the measured junction voltage.  However $R_b$ must be %%@
large enough so that the junction voltage will not retrap immediately after switching, during the hold time.  We find %%@
that it is necessary to increase $R_b$ for smaller switching currents, and that the detailed nature of the junction %%@
I-V curve can play an important role in the choice of $R_b$.  For example, resonances with high frequency %%@
electromagnetic modes of the environment can give parasitic peaks in the junction I-V curve, which inhibit the %%@
latching property of the circuit when $R_b$ is too small.  Such resonant Cooper pair tunneling at finite voltage with %%@
associated excitation of the electromagnetic environment, is most likely the major source of dissipation in this %%@
detector, where quasiparticle tunneling rates are quite suppressed at these temperatures and voltages, well below the %%@
gap energy.  Note that we pull the voltage back to zero before the capacitance of the bias leads is fully charged to %%@
the quiescent operating point of the circuit. However, the value of $R_b$ is chosen such that the junction never %%@
reaches the gap voltage $V_{2\Delta}\approx 400\mu$V and is typically kept below $80\mu$V, as can be seen in %%@
fig.~\ref{measurements}b. 

In order to study how the switching or not-switching of the junction effects subsequent measurements in the pulse %%@
sequence, we need to not only count switching events, but also to store the actual binary sequence of switching in %%@
temporal order.  A fast digital sampling oscilloscope (DSO) with deep memory was used to capture and store the entire %%@
burst sequence of $10^4$ pulses from the AWG, and the amplified sample voltage $U'$.  The data could be analyzed to %%@
reconstruct the binary sequence of switching events.  Correlation functions were studied as a function of the wait %%@
time between pulses, $\tau_w$. 

\section{Results}
\begin{figure}[!ht] 
%\centerline{\epsfxsize=2.7in\epsfbox{DI_over_I.eps}}   
\includegraphics[scale=0.4]{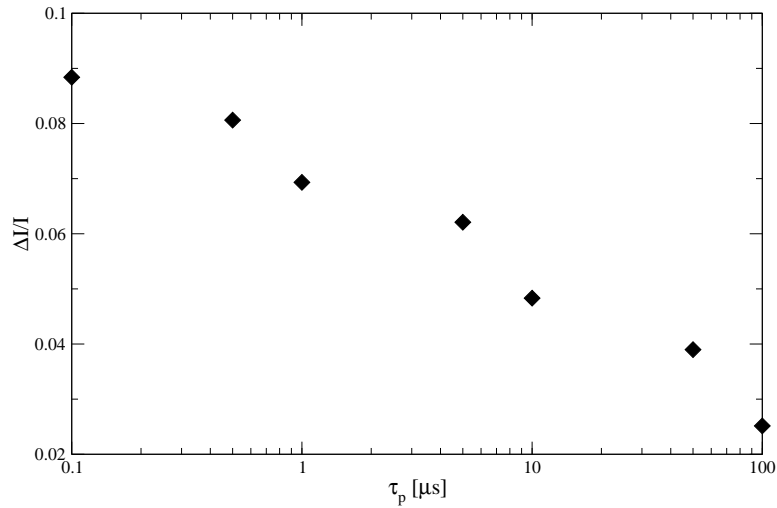}
\caption{Dependence of the resolving power $\Delta I/I$ of the detector as a function of the pulse width $\tau_p$ for %%@
a sample with $I_{switch}\simeq 11$nA.\label{deltaI}}
\end{figure}

The measurement scheme described above can be used to detect the energy eigenstate of a quantum circuit.  For this %%@
purpose it is desirable to have a resolution $\Delta I/I$ which is as narrow as possible, and a measurement time %%@
$\tau_p$ as short as possible. Figure~\ref{deltaI} shows the dependence of $\Delta I/I$ for different values of %%@
$\tau_p$. The magnetic field of the SQUID was tuned such that the critical current for this set of data was calculated %%@
to be $I_{c0}\approx 28nA$.  The switching current however was $\simeq 11$nA. As one can see in fig.~\ref{deltaI}, %%@
even for the short pulses with $\tau_p=100ns$, the  resolving power of the detector is better than $1nA$.  The %%@
increase of $\Delta I/I$ for shorter pulses was observed over the entire range of critical currents measured with this %%@
SQUID.  We also found essentially the same qualitative behavior for a sample with an identical SQUID shunted with a %%@
$1$nF chip capacitor, bonded next to the junction chip, and for a sample where a $10$k$\Omega$ bias resistor bonded %%@
next to the junction chip was used to bias the junction. 

\begin{figure}[!ht] 
%\centerline{\epsfxsize=2.7in\epsfbox{correlation0G.eps}}
\includegraphics[scale=0.4]{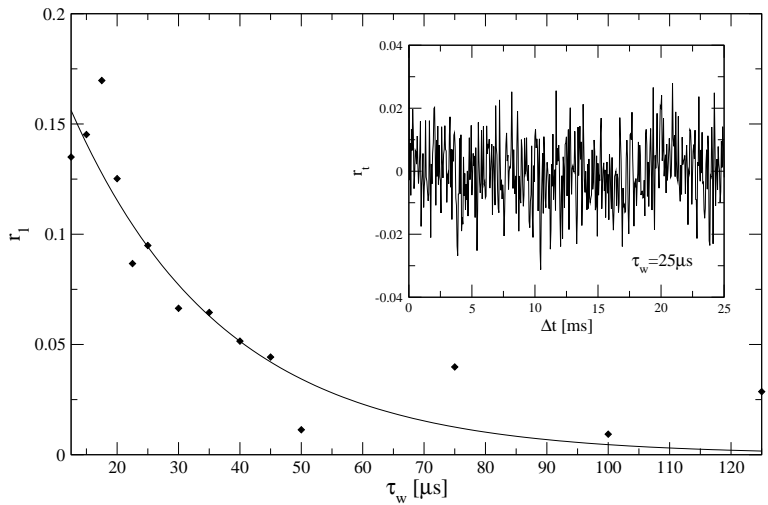}
\caption{Auto correlation for switching events for a sample with $I_{c0}=270$nA where $\tau_p=1\mu$s and %%@
$\tau_h=20\mu$s. The solid line is a fit to the data which gave
$r_1=0.26 exp(-0.04\tau_w)$. The inset shows the correlation as a function of the lag between %%@
pulses.\label{correlation}}
\end{figure}

If the detector is working in an ideal way, each switching event should be statistically independent.  We can check %%@
for this independence by making a correlation analysis of the binary sequence of switching events, ordered as they %%@
were measured.  We captured the entire sequence and analyzed the data, assigning a value $Y_i=0$ for no switch, and %%@
$Y_i=1$ for switch, in chronological order with the index $i$.  We study the auto correlation function of this %%@
sequence which is defined as
\begin{equation} 
r_k=\frac{ \sum_{i=1}^{N_0-k} (Y_i-\bar{Y}) (Y_{i+k}-\bar{Y})} {\sum_{i=1}^{N_0} (Y_i-\bar{Y})^2 }.
\label{autocorr}
\end{equation}
Here, $N_0=10^4$ is the total number of pulses and $\bar{Y}=P(I)$ is the switching probability for the particular %%@
pulse sequence applied.  The  auto correlation, $r_1$, with lag $k=1$, is shown in fig.~\ref{correlation} as a %%@
function of the wait time between measurement pulses, $\tau_w$.  For each data point in fig.~\ref{correlation}, the %%@
pulse was adjusted such that $\bar{Y}=P(I)=0.5 \pm 0.05$, and the data shown is for the case zero magnetic field, or a %%@
large critical current, $I_{c0} = 270$nA.  Here we clearly see an exponential decay of the correlation between %%@
switching events which are nearest-neighbors in time.  The characteristic time for decay of this correlation is $17.2 %%@
\mu$s. Thus, we must program a wait time $\tau_w \gg 17.2\mu$s to avoid measurement errors in our detector.  

One should note that in principle, negative correlations, or anti-correlations are possible as defined by %%@
equation~\ref{autocorr}. Negative correlation would correspond to the a probe pulse that does not switch the junction, %%@
increasing the probability that the probe pulse will result in a switching event.  Negative correlations might result %%@
if, for example, the dissipation during measurement were larger for non-switching than for switching events.  We have %%@
performed circuit simulations of the switching process with P-spice\cite{pspice} by modeling the junction as a voltage %%@
controlled switch, a standard P-spice component.  Depending on the parameters of the simulation, we find that %%@
especially for simple square pulses, it is possible to realize the case where a non-switching event causes more power %%@
dissipation in the circuit, because a high current level is sustained during the entire pulse if the junction does not %%@
switch.  However, we expect positive correlation for the pulse and hold technique, because in this technique we strive %%@
to make the high-current switch pulse as short as possible so that the detector can decohere the system and the %%@
quantum measurement can be as fast as possible.  The power dissipation will predominantly occur during the hold time, %%@
which only is necessary to "get the answer out".

As can be seen, there is a significant correlation between adjacent pulses for wait pulses shorter than $40\mu s$. The %%@
inset of Figure~\ref{correlation} shows the correlation as a function of the lag time $\Delta %%@
t=(\tau_p+\tau_h+\tau_w)k$.  We note that already at lag $k=2$ the correlation function is within the noise level $r_k %%@
\in {-0.02, 0.02}$.  The random nature of $r_k$ vs. $k$ indicates that there is no periodic signal effecting the %%@
switching process.  For smaller switching currents, however, we have noticed the presence of $50$Hz pick-up on the %%@
measurements, which clearly shows up in the correlation function plotted vs. the lag time, $\Delta t$, as a periodic %%@
signal with period $20$ms.

\section{Discussion}

In the DC I-V curves of the junction (not shown here) for lower critical current, we clearly see a finite slope of the %%@
supercurrent branch, due to phase diffusion.  At high frequency ($f>5$MHz), the impedance of our lossy twisted pairs %%@
is essentially $50 \Omega$.  For this shunt impedance, we expect a junction $Q$ factor $Q=\sqrt{2\pi R^2 C I_{c0} / %%@
\Phi_0}=0.12<<1$, which becomes smaller as the critical current of the junction is suppressed.  In this overdamped %%@
limit phase diffusion is expected at finite temperature, or in the presence of external noise %%@
\cite{martinis:phasedifference:87}.   At the lowest critical current,we find that the switching distributions and the %%@
DC I-V curves are essentially unchanged for temperatures less than $400$mK.  Such behavior is consistent with the fact %%@
that we have no low temperature filters protecting the junction.  Thus, we expect that the dynamics of our junction %%@
with such low critical current, has to be described by overdamped classical dynamics of the Josephson phase in the %%@
presence of excess thermal noise. In this case the switching of the circuit is a complicated process of switching %%@
between two dynamical states of the Josephson phase, over a "dissipation barrier" resulting from a frequency dependent %%@
damping of the junction \cite{joyez:josephsoneffect:99}.    The switching occurs when a critical velocity of the phase %%@
particle is reached, and the switching process is between two dynamical states, a slow phase diffusion state, and a %%@
faster damped free-running state.  It is interesting to note that reasonably good resolving power of switching current %%@
detection can be achieved in this overdamped, diffusive regime and an interesting questions arises as to weather or %%@
not a quantum detector can be designed in this regime of switching.  On the one hand fast measurement might be %%@
possible, because we can always increase the switch pulse amplitude in order to more quickly drive the system to it's %%@
critical velocity.  On the other hand, the actual time for the measurement is not so clearly defined in this regime, %%@
and excessive back action from long measurement time would be disadvantageous.  To our knowledge, no theoretical %%@
attempts have been made to describe quantum measurement from a detector which operates in this regime.  Numerical %%@
simulations are in progress to help clarify these issues.

\section*{Acknowledgments}

This research has been supported by the EU program QIPC under the project SQUBIT, and by the Swedish SSF under the %%@
Center for Nano Device Research.  We are grateful to V. Korenivski, D. Vion, I. Chiorescu, M. Devoret, J. Sj\"ostrand, %%@
A. Karlhede and T. H. Hansson for helpful conversations.

%%%%%%%%%%%%%%%%%%%%%%%%%%%%%%%%%%%%%%%%%%%%%%%%%%%%%%%%%%%%%%%%%%%%%%%
% 
%Use this if your figures are put in a subdirectory having the same
%name as the main latex file, ie: 
%
%      ws-procs975x65/procs-fig1.eps      
%      ws-procs975x65/procs-fig2.eps      
%      ws-procs975x65/procs-fig3.eps      
%      ws-procs975x65/procs-fig4.eps      
%      etc.
%
%\begin{figure}[htbp] %ORIGINAL SIZE: width=1.4TRUEIN; height=1.5TRUEIN
%\figurebox{}{}{procf1} %100 percent
%\caption{Labeled tree {\it T}.}
%\end{figure}
%
%%%%%%%%%%%%%%%%%%%%%%%%%%%%%%%%%%%%%%%%%%%%%%%%%%%%%%%%%%%%%%%%%%%%%%%

%\begin{thebibliography}{0}


\begin{thebibliography}{0}

\bibitem{nakamura:CPbox:99}
Y. Nakamura, Y.~A. Pashkin, and J.~S. Tsai, {\it Nature} {\bf 398},  786  (1999).

\bibitem{vion:quantronium:02}
D. Vion {\it et~al.}, {\it Science} {\bf 296},  886  (2002).

\bibitem{martinis:josephsonjunctionqubit:02}
J.~M. Martinis, S. Nam, and J.~A. and C.~Urbina, {\it Phys. Rev. Lett} {\bf 89},
  117901  (2002).

\bibitem{chiorescu:fluxqubit:03}
I. Chiorescu, Y. Nakamura, C.~J. P.~M. Harmans, and J.~E. Mooij, {\it Science} {\bf
  299},  1869  (2003).

\bibitem{duty:chargequbit:03}
T. Duty {\it et~al.}, {\it Phys. Rev. B} {\bf 69},  140503  (2003).

\bibitem{voss:mqt:81}
R.~F. Voss and R.~A. Webb, {\it Phys. Rev. Lett.} {\bf 47},  265  (1981).

\bibitem{devoret:quantumeffects:92}
M.~H. Devoret {\it et~al.},  in {\em Quantum Tunnelling in Condensed Media},
  edited by Y. Kagan and A.~J. Leggett (Elsevier Science, Amsterdam, 1992),
  Chap.~6, pp.\ 313--345.

\bibitem{fulton:zerovoltagelifetime:73}
T.~A. Fulton and L.~N. Dunkleberger, {\it Phys. Rev. B} {\bf 9},  4760  (1973).

\bibitem{cottet:pulsereadout:02}
A. Cottet {\it et~al.}, {\it Physica C} {\bf 367},  197  (2002).

\bibitem{cottet:CPTsingleshot:02}
A. Cottet {\it et~al.},  in {\em International Workshop on Superconducting
  Nano-electronic Devices}, edited by J. Pekola, B. Ruggiero, and P.
  Silvestrini (Kluwer Academic, Plenum Publishers, New York, 2002), pp.\
  73--86.

\bibitem{vion:privatecomm}
D. Vion, Private communication.

\bibitem{koch:magnetizationreversal:98}
R.~H. Koch {\it et~al.}, {\it Phys. Rev. Lett.} {\bf 81},  4512  (1998).

\bibitem{pspice}
Available free on the internet at www.orcad.com.

\bibitem{martinis:phasedifference:87}
J.~M. Martinis, M.~H. Devoret, and J. Clarke, {\it Phys. Rev. B} {\bf 35},  4682
  (1987).

\bibitem{joyez:josephsoneffect:99}
P. Joyez {\it et~al.}, {\it Journal of Superconductivity} {\bf 12},  757  (1999).

\end{thebibliography}
\end{document}